\documentstyle[12pt,particles,journals,symbols,epsfig]{article}
\def\ngspace{\kern -.04em}

\def\vrnm{\vrule height1.00ex depth0.55ex width0em}
\def\vrdn{\vrule height1.55ex depth0.00ex width0em}
\def\hrwa{\vrule height0pt depth0pt width0.10em}
\def\hrwb{\vrule height0pt depth0pt width0.25em}
\def\pmth#1#2#3{#1%
     {\textstyle\vrnm\hrwa+\hrwb#2\atop\textstyle\vrdn\hrwa-\hrwb#3}}
\def\pmfv#1#2#3#4#5{#1%
     {\textstyle\vrnm\hrwa+\hrwb#2+#4\atop\textstyle\vrdn\hrwa-\hrwb#3-#5}}    
\def\vrht{\vrule height2.6ex depth0pt width0pt}
\catcode`@=11
\def\eqalign#1{\null\,\vcenter{\openup\jot\m@th
  \ialign{\strut\hfil$\displaystyle{##}$&$\displaystyle{{}##}$\hfil
      \crcr#1\crcr}}\,}
\catcode`@=12
\def\bln#1#2\eln{\begin{equation}\label{#1}%
   \eqalign{#2}\end{equation}\vskip3mm\noindent}

\def\brlist{\begin{thebibliography}{99}}
\def\brf{\bibitem}
\def\erlist{\end{thebibliography}}
\def\mcite{$\,$\cite}

\def\brlist{\begin{thebibliography}{99}}
\def\brf{\bibitem}
\def\erlist{\end{thebibliography}}
\begin{document}
\setlength{\baselineskip} {2.5ex}
\hyphenation {che-ren-kov}
\hyphenation {non-per-tur-ba-tive}
\hyphenation {per-tur-ba-tive}
\begin{center}
\vspace{3.cm}
{Contributions to the International Workshop on Hadron Physics\\ with High
Energy Muon and Hadron Beams at Fixed Target Experiments,\\ Workshop
Chairmen S. Paul and F. Bradamante, Technical University Munich, Oct.
1999,\\ Tel Aviv U. Preprint
TAUP-2605-99\\}
\vspace{0.6in}
{\Large\bf
Hybrid Meson Structure at COMPASS \\}
\vspace{0.5cm}
{\Large
Murray Moinester$^{a}$, 
Suh Urk Chung$^{b}$\\}
\end{center}
\vspace{0.2cm}
{\bf \em a--School of Physics and Astronomy, R. and B. Sackler Faculty of
Exact 
Sciences,\\ Tel Aviv University, 69978
Tel Aviv, Israel, e-mail: murraym$@$tauphy.tau.ac.il 
\\b--Department of Physics, Brookhaven National
Laboratory,\\ Upton, NY 11973, USA, e-mail: suchung@qgs.phy.bnl.gov\\}
\vspace{1cm}
\begin{center}
{\large\bf  Abstract}
\end{center}
\vspace{.2cm}
{\bf Objectives and Significance:\\}

We describe a pion physics program attainable with the CERN COMPASS
spectrometer, involving tracking detectors and an electromagnetic
calorimeter. COMPASS can realize state-of-the-art pion beam hybrid meson
and meson radiative transition studies. We review here the physics
motivation for this program. We describe the beam, detector, trigger
requirements, and hardware/software requirements for this program. The
triggers for all this hybrid meson physics can be implemented for
simultaneous data taking. 

\noindent 

We will investigate hybrid meson production via pion-photon Primakoff and
pion-Pomeron diffractive interactions. We will determine new properties of
quark-antiquark-gluon hybrid mesons, using unique production methods, to
improve our understanding of these exotic mesons.  \\
 
\noindent
{\bf Methodology:\\}
\noindent 

The CERN COMPASS experiment uses 100-280 GeV beams ($\mu$, $\pi$), and
magnetic spectrometers and calorimeters, to measure the complete
kinematics of pion-photon and pion-Pomeron reactions. The COMPASS
experiment is currently under construction, and scheduled to begin data
runs in 2001.  We carry out simulation studies to optimize the beam,
detector, trigger, and hardware/software for achieving high statistics
data with low systematic uncertainties in the hybrid meson component of
this program. We will improve previous Primakoff Hybrid studies by three
orders of magnitude. We implement special detectors and triggers for
hybrid meson production reactions.  We propose to prepare for these
COMPASS pion beam hybrid studies by setting up with muon beam tests.

\newpage
\begin{center}
{\bf  1. Description of Subject:\\}
\end{center}

\noindent 

The COMPASS physics programs \cite {compass} include studies of
pion-photon Primakoff and pion-Pomeron diffractive interactions using
100-280 GeV/$c$ negative pion beams in dedicated data runs.  Hybrid mesons
can be studied in this way, and can provide significant tests of QCD
predictions.  COMPASS Primakoff planning studies were described in recent
workshop proceedings \cite {cd2,hadron1,bormio}, and at COMPASS
collaboration meetings. 

\vspace{.5cm}
\noindent
{\bf 1.1~Hybrid Mesons\\}

\noindent

The hybrid ($q\bar{q}g$) mesons, along with glueballs ($gg$) are some of
the most interesting consequences of the non-Abelian nature of QCD.
Detection of these exotic states is a long-standing experimental puzzle. 
The most popular approach for hybrid searches is to look for the
`oddballs'---mesons with quantum numbers not allowed for ordinary
$q\bar{q}$ states.  For Primakoff/diffractive production, the outgoing
mesons preserve the charge of the incoming beam, i.e. $I=1$ for the
resonances under study.  Then, the `oddball' mesons for $J\leq 2$ come in
the following variety: 

\begin{center}
\begin{tabular}{c|c}$I^G(J^{PC})$&`Oddball'\\ \hline $1^+(0^{--})$&$\rho_0$\\
$1^+(0^{+-})$&$b_0$\\ $1^-(1^{-+})$&$\pi_1$\\ $1^+(2^{+-})$&$b_2$\\
\end{tabular}
\end{center} 
\vskip2mm

Barnes and Isgur first discussed hybrid meson properties, and more
recently in the flux-tube model \cite {tb,ni}. In the flux-tube model, the
mass of the lightest gluonic hybrid is predicted be around 1.9 GeV, with
the quantum numbers of $J^{PC}=1^{-+}$.  Close and Page \mcite{fluxtb}
predict that such a gluonic hybrid should decay into the following
channels: 

\begin{center}\begin{tabular}{c|c|c|c|c}
   $b_1\pi$&$f_1\pi$&$\rho\pi$&$\eta\pi$&$\eta'\pi$\\
\hline
170&60&$5\to 20$&$0\to 10$&$0\to 10$\\
\end{tabular}\end{center}
\vskip2mm

\noindent
where the numbers refer to the partial widths in MeV.  According to them,
its total width must be larger than 235-270 MeV, since the $s+\bar s$
decay modes were not included.  Recent updates on hybrid meson structure
are given in Refs. \cite{nib98,kpb98,sgjn}


   From more than a decade of experimental efforts at IHEP \cite
{ihep1,ihep2,ves}, CERN \cite {na12}, KEK \cite{kek}, and BNL \cite
{E852}, several hybrid candidates have been identified. More recently, new
information came from the BNL E852 experiment \cite {E852}, which studied
the $\pi^- p$ interaction at 18 GeV/$c$. They reported two $J^{PC}=
1^{-+}$ resonant signals at masses of 1.4 and 1.6 GeV in $\eta\pi^-$ and
$\eta\pi^0$ systems, as well as in $\pi^+ \pi^- \pi^-$, $\pi^- \pi^0
\pi^0$, $\eta' \pi^-$ and $f_1(1285)  \pi^-$. Also, a VES group \cite
{ves} has published analyses of $\eta \pi^-$, $\eta' \pi^-$, $f_1(1285) 
\pi^-$, $b_1(1235)  \pi^-$ and $\rho \pi^-$ systems produced in $\pi^-
{\rm Be}$ interactions at 37 GeV/$c$.  VES sees the $J^{PC}= 1^{-+}$ wave
clearly in all channels, and they report an indication of a resonance at
1.6 GeV.  It is striking that the VES phase motion of the $1^{-+}$ wave in
$\eta\pi^-$ shows a rise at 1.4 GeV, identical to that of the BNL E852
data.  The most recent information on the 1.4-GeV state comes from two
analyses by the Crystal Barrel collaboration on the $\bar p p$ and $\bar p
n$ annihilations at rest into $\pi\pi\eta$ \cite{CBa},\cite{CBb}.  Their
observed masses and widths are consistent with those of BNL E852.  VES
reports that the ratio of $\eta' \pi$ to $\eta\pi$ $P$-waves at 1.4 GeV is
low, while that at 1.6 GeV is high. This is considered as evidence that
the hybrid nature of the exotic wave at 1.6 GeV is gluonic; i.e., its
constituents are $q+\bar q+{\rm gluon}$, where $q$ stands for light
nonstrange quarks. 

   The partial-wave analysis (PWA) of systems such as $\eta\pi$ or $\eta'
\pi$ in the mass region below 2 GeV requires care and experience. This is
so because (1) this region is dominated by the strong $2^+$ `background'
($a_2$ resonance), and (2) that the PWA may give ambiguous results \cite
{ihep2} for the weaker $1^{-+}$ wave. For Primakoff production, the hybrid
production cross section may increase relative to the $a_2$ state,
considering the estimated radiative widths. These are $\Gamma(a_2
\rightarrow \pi \gamma)=300$~ke$\!$V, and $\Gamma(\pi_1 \rightarrow \pi
\gamma)\approx 90-540$~ke$\!$V, as discussed in Section 2.  Therefore, the
PWA uncertainties for the $1^{-+}$ wave will be different and may even
improve. The problem generally is that the PWA of the $\eta\pi$ system
must take into account $S$-, $P$- and $D$-waves, and the number of
observables is not sufficient to solve all equations unambiguously.  The
strength and phase ambiguities as a function of mass of different partial
wave solutions are discussed in ref. \cite {ihep2}. However, in certain
experimental situations, the ambiguous solutions have relatively little
impact on a particular exotic wave under study---such a situation seems to
be the case with the BNL E852 data.  In addition, it has been shown
\cite{E852rev} that certain other assumptions, e.g., the rank-one
condition, can be removed in a systematic study in which the
mass-dependence of each partial wave is introduced explicitly into the
analysis. 

   The masses and widths of $\pi_1(1400)$ meson in the decay channel
$\pi\eta$ are summarized in Table I. 

\def\arraystretch{1.8}
\begin{center}\begin{tabular}{|l@{\hspace{2mm}}|l@{\hspace{2mm}}|l|}
\multicolumn{3}{c}{Table I: Parameters for $\pi_1(1400)\to \eta\pi$}\\
\hline\hline
Expt. & Mass(MeV) & Width(MeV)\\
\hline
KEK & $1323.1\pm 4.6\vrht $ & $143.2\pm 12.5$\\
BNL('94) & $1370\pm\pmth{16}{50}{30}$ 
              & $385\pm\pmth{40}{\phantom{1}65}{105}$\\
BNL('95) & $\pmfv{1359}{16}{14}{10}{24}$ 
            & $\pmfv{314}{31}{29}{\phantom{1}9}{66}$\\
CB &$1400\pm 20\pm 20$ & $310\pm\pmth{50}{50}{30}$\\
CB &$1360\pm 25$ & $220\pm 90$\\
\hline
\end{tabular}\end{center}
\vskip2mm

At a recent Workshop on Hadron Spectroscopy \cite {whs}, the VES
collaboration presented the results of a coupled-channel analysis of the
$\pi_1(1600)$ meson in the channels $\rho\pi$, $\eta'\pi$ and
$b_1(1235)\pi$. Their results are consistent with the BNL results, as seen
in Table II. 

\begin{center}\begin{tabular}{|l|l|l|l|}
\multicolumn{4}{c}{Table II: Parameters for $\pi_1(1600)$ Decay}\\
\hline\hline
Expt.& Mass(MeV) & Width(MeV)& Decay\\
\hline
BNL  & $1593\pm\pmth{8}{20}{47}$ 
        & $168\pm \pmth{20}{150}{\phantom{1}12}$&$\rho\pi$\\
BNL  & $1596\pm 8$ & $387\pm 23$&$\eta'\pi$\\
VES &$1610\pm 20$ & $290\pm 30$&$\rho\pi$,$\eta'\pi$,$b_1\pi$\\
\hline
\end{tabular}\end{center}
\vskip2mm


   For both BNL E852 and VES data, it is not known what Regge exchanges
are responsible for the production of the $J^{PC}=1^{-+}$ exotic states at
1.4 and 1.6 GeV, Both the $a_2(1320)$ and the exotic waves are produced
via natural-parity exchanges which include the Pomeron.  If Pomeron
exchange is indeed responsible the production, then diffractive production
in COMPASS can provide an additional handle with which to tackle the study
of exotic waves.

   One can succinctly summarize the situation as follows: a production of
the wave $I^G(J^{PC})=1^-(1^{-+})$ is dependent on the strength of the
$\pi\rho$ decay modes in the case of the Primakoff production, whereas in
diffractive production the relative strengths depend on the supposed decay
modes $\pi_1(1400)\pi$ and $\pi_1(1600)\pi$ of the tensor glueball
(2$^{++}$), since the Pomeron is thought to be on the Regge trajectory
corresponding to the tensor glueball with a presumed mass around 2 GeV.
Corresponding to the glueball decay $G(2^{++}) \rightarrow
\pi^+ Hybrid$,
one expects diffractive production via $\pi^- G(2^{++})
\rightarrow
Hybrid$.  
This is an additional strong advantage of the COMPASS hybrid meson study. 
We can look forward to two complementary production modes of exotic
mesons, increasing our chance for achieving a decisive advance on our
understanding of the meson constituents.  Finally, the E852 collaboration
finds preliminary evidence of a third exotic meson at around 1.9 GeV. The
search for this state as well as others can continue, with exciting new
results anticipated.  In summary, COMPASS can move into a forefront of
hadron spectroscopy, by studying Primakoff and diffractive production of
nonstrange light-quark hybrid mesons in the 1.4-2.5 GeV mass region,
including all the hybrid candidates from previous studies.

\vskip5mm
\noindent
{\bf 1.2~Radiative Transitions\\}

\noindent

Radiative decay widths of mesons and baryons are powerful tools for
understanding the structure of elementary particles and for constructing
dynamical theories of hadronic systems. Straightforward predictions for
radiative widths make possible the direct comparison of experiment and
theory.  The small value of branching ratios of radiative decays makes
them difficult to measure directly, because of the large background decay
$\pi^0$s from strong decays.  Studying the inverse reaction
\mbox{$\gamma+\pi^-\rightarrow M^-$} provides a relatively clean method
for the determination of the radiative widths. Very good tracking
resolution is needed (and available through silicon strip detectors) to
measure initial and final state momenta, and to thus exhibit the Primakoff
signal at small momentum transfers, where the electromagnetic processes
dominate over the strong interaction. 

In COMPASS, we will study radiative transitions of incident mesons to
higher excited states.  We will obtain new data (\cite {hadron1})  for
radiative transitions leading from the pion to a$_1$(1260), a$_2$(1320),
and $\rho$ mesons. The previous Coulomb field measurements of the
a$_1$(1260)  $\rightarrow \pi \gamma$ width (\cite{ziel}) is $0.64 \pm
0.25 $ MeV;  of the a$_2$(1320)  $\rightarrow \pi \gamma$ width
\cite{ciha} is 0.30 $\pm$ 0.06 MeV; and of the $\rho \rightarrow \pi
\gamma$ width (\cite{jens}) is 60 to 81 ke$\!$V. We will obtain independent
and significantly higher precision data and statistics for these and
higher resonances.  This will be valuable in order to allow more
meaningful comparisons with theoretical predictions, and as a
normalization of the Hybrid meson studies.

\vspace{.5cm}
\noindent
{\bf 1.3~Experimental Requirements\\}

We consider the beam, target, detector, and trigger requirements for
hybrid meson production and detection with minimum background
contamination. 

\vspace{.5cm}
\noindent
{\bf 1.3.1~Monte Carlo Simulations\\}

We carry out Monte Carlo simulations with HYBRID, an event generator
adapted to COMPASS for Hybrid meson physics studies.  For hybrid mesons,
simulations \cite {zihy} were carried out for the FNAL SELEX apparatus, a
low rate forward spectrometer, otherwise very similar to COMPASS. SELEX
however did not obtain quality hybrid physics data. We base our initial
planning on the previous FNAL Primakoff experiments and on available SELEX
simulations, while pursuing further simulations for COMPASS \cite {mchy}. 

\vspace{.5cm}
\noindent
{\bf 1.3.2~Beam Requirements\\}

A beam Cherenkov detector (CEDARS) far upstream of the target provides
$\pi/K/p$ identification.  We will take data with both positive and
negative beams. We may use a 280 GeV beam, the highest energy available at
COMPASS, because Hybrid meson Primakoff production cross sections increase
with increasing energy, and because the calorimeter acceptances are higher
at the highest energy. Beam rates lower than the 40 MHz COMPASS design
rate are planned for initial setup studies, in which many of the COMPASS
systems (DAQ, detectors, etc.) must be implemented.  We base our count
rate estimates on beam intensities of 10 MHz ($20 \times 10^7$ particles
in a 2 second spill with a total cycle time of 14.8 seconds). \\

\vspace{.5cm}
\noindent
{\bf 1.3.3~Target and Target Detectors\\}

We mainly veto target break-up events by positioning veto scintillators
around the target. We also veto target break-up events by selecting
multiplicity 1 or 3 events in downstream hodoscopes H1 and H2 (Fig. 2) at
the trigger level, and by selecting low-t events in the off-line analysis.
Before and after the target, charged particles are tracked by high
resolution tracking detectors.  We achieve good angular resolution for the
final state charged particles by minimizing the multiple scattering in the
targets and detectors.  The Primakoff targets will be Pb of 1\%
interaction length = 2 g/cm$^2$ =0.30 radiation length, and other targets
such as Cu of similar and also smaller radiation length.  The beam and
outgoing pion multiple Coulomb scattering in the target gives an rms
angular resolution of about 40 $\mu$rad.  \\

\vspace{.5cm}
\noindent
{\bf 1.3.4~The Magnetic Spectrometer and the t-Resolution\\}

The incoming beam momentum is measured with upstream SPS detectors. The
final state pion and $\gamma$ momenta are measured with good resolution in
downstream COMPASS magnetic spectrometers and in the photon calorimeter,
respectively.  Via the measurement of incident and final state momenta, we
obtain a precise determination of the square of the four momentum transfer
${\it t}$ to the target nucleus.  The small transverse momentum kick p$_T$
to the target and ${\it t}$ are related by ${\it t} = p_T^2$.  We aim for
a p$_T$ resolution or about 10 MeV/c, corresponding to resolution $\Delta
t = 2 \times 10^{-4}$ GeV$^2$ for production of a 2 GeV Hybrid.  This goal
is based on the need to minimize contributions to the Coulomb Primakoff
data from diffractive production. A peak in the ${\it t}$ distribution at
low ${\it t}$ provides the main signature of the Primakoff process, and
the means to separate Primakoff from diffractive scattering. The Pb
diffractive data for example falls as $\exp(-t/0.0025)$ with t expressed
in GeV$^2$. Our t-resolution goal is then about a factor of 10 smaller
than the slope in t observed for diffractive data on a Pb target. This
goal is clearly achievable, as one may see from the t distributions
measured at a low statistics but high resolution experiment for $\pi^-
\rightarrow \pi^- \pi^0$ \cite {jens} and $\pi^- \rightarrow \pi^- \gamma$
\cite {pigam} Primakoff scattering at 200 GeV at FNAL. The t distribution
of the $\pi^- \rightarrow \pi^- \gamma$ \cite {pigam} data agrees well
with the Primakoff formalism out to t~=~$10^{-3}$ GeV$^2$, which indicates
that the data are indeed dominated by Coulomb production.  \\

\vspace{.5cm}
\noindent
{\bf 1.3.5~The Photon Calorimeter ECAL2\\}

The momentum kicks of the two COMPASS magnets are set $\it{additive}$ for
maximum deflection of the beam from the zero degree (neutral ray)  line. 
This maximizes the distance from the zero degree line to the beam hole in
ECAL2 (located about 30 meters from target), and attains an acceptable
distance of at least 10 cm between the zero degree line and the deflected
beam position at ECAL2 for the proposed 280 GeV beam energy. The hole size
and position must be optimized to minimize the number of pions hitting
ECAL2 blocks at the hole perimeter. The Primakoff $\gamma$'s are centered
around the zero degree line, and a good $\gamma$ measurement requires
clean signals from 9 blocks, centered on the hit block. The ADC electronic
readouts for these blocks have been designed and are being built. 

As can be seen in Fig. 1, COMPASS needs to also detect $\eta$s for the
hybrid study.  The two $\gamma$s from $\eta$ decay have half-opening
angles $\theta_{\gamma\gamma}^h$ for the symmetric decays of
$\theta_{\gamma\gamma}^h= m/E_{\eta}$, where m is the mass ($\eta$)  and
E$_{\eta}$ is the $\eta$ energy. Opening angles are somewhat larger for
asymmetric decays. In order to catch about 50\% of the decays, it is
necessary to subtend a cone with double that angle, i.e.  $\pm
2m/E_{\eta}$, neglecting the angular spread of the original $\eta$s around
the beam direction.  Consider an ECAL2 $\gamma$ detector with a circular
active area with 2 m diameter.  Consider the $\pi\eta$ channel.  For an
ECAL2 of 1 m radius at 30 m from the target, $\eta$s above E$_{\eta}$=33
GeV are therefore accepted.  At half this energy, the acceptance
practically vanishes. The acceptance of course depends on the Hybrid mass,
mostly between 1.4 and 2.5 GeV for the planned COMPASS study. Detailed
Monte Carlo studies are in progress for the different possibly Hybrid
decay modes, for a range of assumed masses.  One also must consider the
efficiency that the two $\gamma$s are sufficiently separated, to be able
to get a position and energy measurement for each of them. For the
$\pi$f$_1$ channel, with for example $f_1 \rightarrow \pi\pi\eta$, the
$\eta$s will have low energy, and therefore large gamma angles. To
maintain good acceptance for low energy $\eta$s, the ECAL2 diameter should
be 2 m or more.

Primakoff physics requires a very good position and energy resolution of
photon calorimeters. The ECAL2 blocks will have their gains well matched,
and their analog signals will be electronically summed and discriminated
to provide a trigger signal on minimal energy deposit.  For the precise
monitoring of the energy calibration of the photon calorimeters, COMPASS
will use a dedicated laser system ~\cite {h741}. 

\vspace{.5cm}
\noindent
{\bf 1.3.6~The Primakoff Trigger\\}

We design \cite {cd2,hadron1,bormio} the COMPASS Primakoff/Diffractive
hybrid meson trigger to enhance the acceptance and statistics, and also to
yield a trigger rate closer to the natural rate given by the low hybrid
cross sections.  The trigger should suppress the beam rate by a factor
10$^3$-10$^4$ or better, while also achieving high acceptance.  The
resulting rate will be significantly lower than the maximum of 10$^5$ per
second DAQ limit in COMPASS \cite {daq}.  We veto target break-up events
via veto scintillators around the target. For hybrid meson physics, the
trigger uses the characteristic hybrid decay pattern: one or three charged
hadrons with gamma hits, or three charged hadrons and no gamma hits.  The
hybrid trigger \cite {cd2,hadron1,bormio} for the $\pi\eta$ hybrid decay
channel (charged particle multiplicity =1) is based on a determination of
the pion energy loss (via its characteristic angular deflection),
correlated in downstream scintillator hodoscopes stations (H1 versus H2) 
with the aid of a fast matrix chip, as shown in Fig. 2. The chip was
designed and already tested for the analogous COMPASS energy-loss trigger
for the muon beam runs. 

We will also test alternative and/or complementary trigger concepts.  We
have already successfully carried out trigger tests at the CERN test
channels, and made reports at COMPASS collaboration meetings.  For
example, the non-interacting beam may be detected and vetoed by the Beam
Kill veto trigger detectors BK1/BK2, which follow the pion trajectory, as
shown in Fig. 2.  This Beam Kill idea must be tested. It requires very
fine segmentation for the BK1 and BK2 detectors to be able to accept a 40
MHz beam rate, and it requires very high efficiency since detector
inefficiencies in killing the beam can lead to artificially high trigger
rates. We will test the energy loss trigger described above, with and
without adding the beam kill trigger capability. Depending on the results,
we will make a decision on implementing the BK detectors. Even if finally
not implemented for the data taking, tests with the beam killer detectors
will be valuable for setting up the energy loss trigger. 

We foresee a three-level Primakoff trigger scheme: T0 = beam definition,
T1 = event topology, T2 = online software filter. T0 is a fast logical
signal that includes upstream CEDARS Cherenkov detectors for beam PID. T1
is a downstream coincidence between scintillation H1-H2 hodoscope signals,
with charged particle multiplicity 1 ($\pi^-$) or 3 (2 negative, one
positive) conditions.  T2 is an intelligent software filter, placed in the
DAQ stream after the event builder, which counts the number of
reconstructed track segments upstream and downstream from the target, and
also sets cuts on event quality characteristics (goodness of kink or
vertex reconstruction, cuts on the t-distribution, etc.). 

\begin{figure}[tbc]
\centerline{\epsfig{file=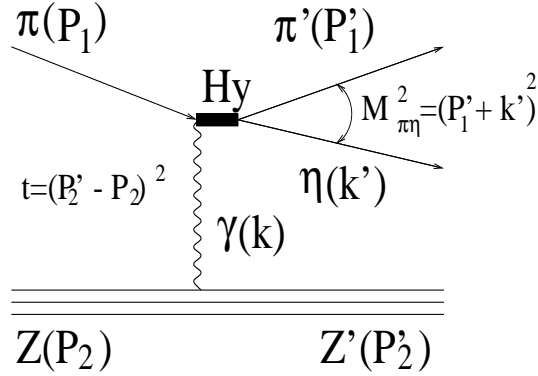,width=7cm,height=5cm}}
\caption{The Primakoff $\gamma$-pion Hybrid production process and
kinematic variables (4-momenta): P1, P1$^\prime$ = for initial/final pion,
P2, P2$^\prime$ = for initial/final target, k = for initial $\gamma$,
k$^\prime$ = for final $\eta$.}
\label{fig:diagram}
\end{figure}

As an example, for M$_{HY}$ near 1.5 GeV, HY $\rightarrow \pi^- \eta$
events can have 40-235 GeV pions at angles larger than 0.5 mrad, close to
the non-interacting 280 GeV beam, and two forward photons at angles less
than 30 mrad. Pions with energy lower than 40 GeV are blocked by the
magnet yokes. The kinematic variables for the HY $\rightarrow \pi^- \eta$
Primakoff process are shown in Fig.~\ref{fig:diagram}. A virtual photon
from the Coulomb field of the target nucleus interacts with the pion, a
Hybrid meson is produced and decays to $\pi^- \eta$ at small forward
angles in the laboratory frame, while the target nucleus (in the ground
state)  recoils coherently with a small transverse kick p$_T$. The peak at
small target $p_T$ used to identify the Primakoff process is measured
offline using the beam and vertex silicon detectors. For diffractive
processes, the beam pion interacts with an exchanged Pomeron.

\begin{figure}[tbc] 
\centerline{\epsfig{file=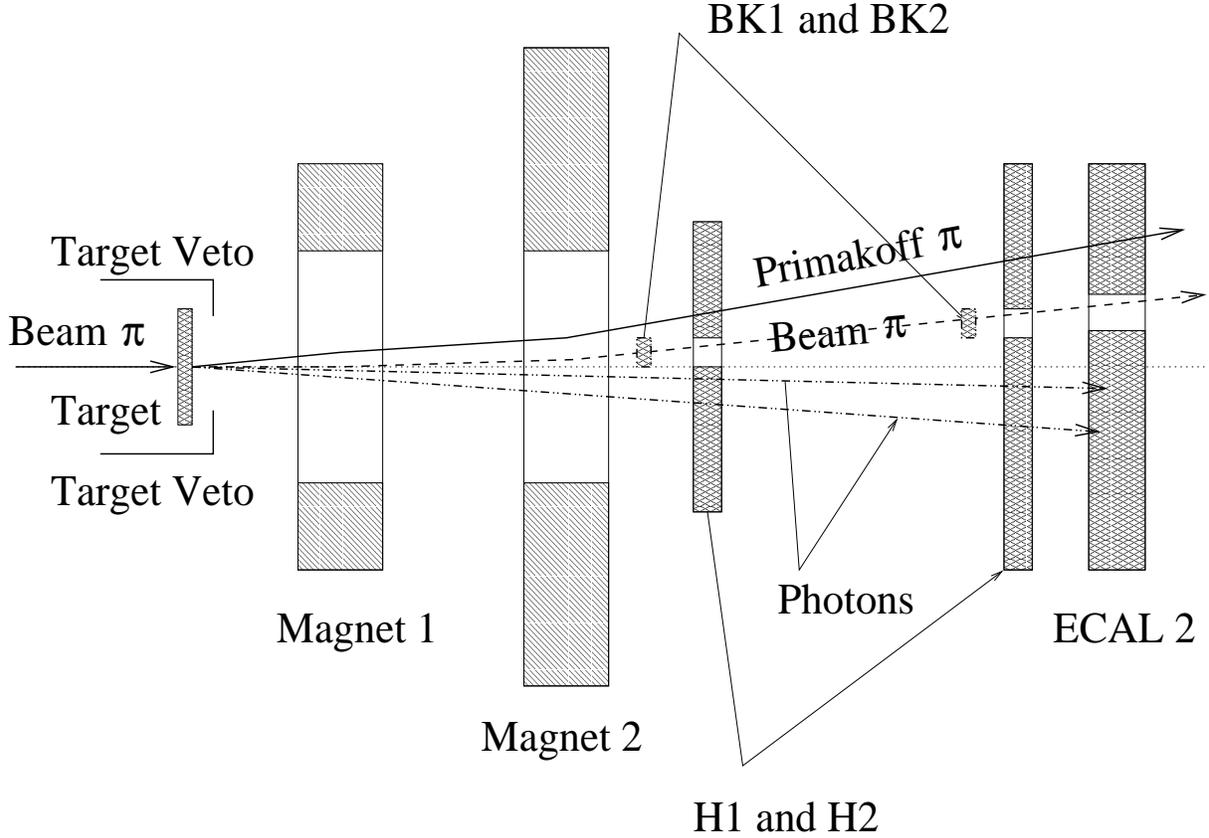,width=16cm,height=11cm}}
\caption{Detector layout for the COMPASS Primakoff Hybrid trigger.
BK1,BK2=beam killer system, H1,H2=hodoscope system for charged particle 
detection, ECAL2=second photon calorimeter.}
\label{fig:trigger}
\end{figure}

The T1 trigger scheme (for photon-pion reactions with one charged pion and
two photons in the final state) for an assumed 280 GeV pion beam is shown
in Fig.~\ref{fig:trigger}. BK1 and BK2 are small scintillator hodoscopes; 
while H1 and H2 are larger scintillator hodoscopes with larger
segmentation, all in the beam bend plane. The beam passes through holes in
H1 and H2.  An anticoincidence in the ``beam" region of BK1 and BK2 can be
used to veto non-interacting beam pions. According to Monte Carlo
simulation, a 30 GeV energy loss condition achieves beam suppression with
99.8\% efficiency, while maintaining 100\% efficiency for all Primakoff
scattered pions with momenta at least 30 GeV/c lower than the beam
momentum. For a trigger that accepts Primakoff scattered pions with
momenta at least 45 GeV/c lower than the beam momentum, beam suppression
is of course yet better. For the case of one charged pion in the final
state, we require an ECAL2 $\gamma$ signal above 45 GeV in coincidence
with a final-state pion in the energy range $(40-235~ \rm{GeV})$.  The
threshold for the ECAL2 $\gamma$ signal is set to match the kinematic
region of Primakoff scattered pions that satisfy the 45 GeV energy loss
trigger condition.  The BK1 and BK2 hodoscope sizes are optimized for beam
pions. We measure the energy loss of the $\pi^-$ via characteristic
angular deflections, correlated with hits in the H1 versus H2 hodoscopes.
A fast matrix chip is used for this purpose, as developed for the standard
muon energy-loss trigger in planned COMPASS studies of gluon polarization
in the proton.  A coincidence based on H1/H2 hodoscope correlations, with
multiplicity=1 condition, where the H1-H2 line projects back to the target
beam spot, is used to trigger on the Primakoff decay pion in the
$\pi^-\eta$ channel.

This trigger configuration for Primakoff scattering strongly suppresses
backgrounds associated with both non-interacting beam particles and those
involving nuclear interaction of pions in the target and in COMPASS
apparatus. The Primakoff trigger design suppresses the beam rate by up to
a factor 10$^3$-10$^4$, achieving high acceptance efficiency for events
versus the important kinematic variables. The diffractive cross sections
leading to the $\pi^-\pi^0\pi^0$ and $\pi^-\pi^-\pi^+$ final states are
large.  Since we aim to also observe diffractive production, we may need
to prescale the triggers for such strong channels.

We will study the acceptance of this trigger in COMPASS for the
$\pi^-\eta$ hybrid meson decay mode using our MC code HYBRID \cite {mchy},
which generates Primakoff pion-photon hybrid meson production
interactions, with realistic beam phase space. We will also study the $\pi
f_1$ and other decay mode acceptances with HYBRID. The T1 trigger scheme
for the $\pi f_1$ decay channel is based on part or all of the following:
a multiplicity = 3 condition, condition of one positive and 2 negative
tracks, condition that one can find 3 H1-H2 track lines that all point
back to the small beam spot on target, condition that the total energy
associated with the tracks and $\gamma$ energy is of order 280 GeV. How
many of these conditions we include depends on how low a trigger rate we
may easily achieve with minimum bias. For the case of three charged
particles and also $\gamma$s in the final state, the trigger rate is lower
than the beam rate by a lower factor than for $\pi\eta$, but still lower
than the COMPASS DAQ limit.  In addition, the H1 and H2 hodoscope may veto
charged particles with larger angles than expected for all the hybrid
decay channels, and also events with multiplicity higher than 3. 

Minimum material (radiation and interaction lengths) in COMPASS will give
a clean (low background) trigger. This is so since $\gamma$s may arrive at
ECAL2 with minimum interaction losses, while producing minimum background
interactions. That is, it is best to use a minimally instrumented COMPASS
apparatus for Primakoff and Diffractive hybrid meson structure studies.

\begin{center}
{\bf  2. Objectives and Expected Significance\\}
\end{center}

\noindent

COMPASS can contribute significantly to the further investigation of
hybrids by studying Primakoff and Diffractive production of
$I^G(J^{PC})=1^-(1^{-+})$ `$\pi_1$'---or more generally
$I^G(J^{PC})=1^+(0^{+-})$ `$b_0$' or $I^G(J^{PC})=1^+(2^{+-})$
`$b_2$'---hybrids.  The possibilities for Primakoff production of the
$\pi_1$ with energetic pion beams, and detection via different decay
channels, were discussed previously in Refs. \cite {bormio,zihy}, and
Monte Carlo simulations for this physics are in progress for COMPASS \cite
{mchy}. The experiment will be run with the COMPASS spectrometer,
consisting of the spectrometer magnets, the central tracking detectors,
the ECAL2 calorimeter, and a relatively simple trigger.  The COMPASS
Primakoff trigger will allow observation of the $\pi_1$ via the
$\eta\pi^-$ decay mode. With a relative $P$-wave (L=1), the $\eta\pi^-$
system has $J^{PC}= 1^{-+}$. The other decay channels of $\pi_1$ may be
studied simultaneously in COMPASS by relatively simple particle
multiplicity triggers (three charged particles in final state, etc.).

We make rough estimates of the statistics attainable for hybrid production
in the COMPASS experiment. Monte Carlo simulations in progress will refine
these estimates. We assume a 125-1250$\mu$b Hybrid meson production cross
section per Pb nucleus (near 1.5 GeV mass). This estimate is based on two
considerations. First, a straightforward application of VDM with
$\rho-\gamma$ coupling g$_{\rho\gamma}^2/\pi$=2.5, gives a width of
$\Gamma(\pi_1 \rightarrow \pi \gamma$) = 75-750 keV for a 1.5 GeV
Hybrid,
assuming $\Gamma(\pi_1 \rightarrow \pi \rho$)= 10-100 MeV, 
a range corresponding to 
3.3-33\%
of the claimed 1.5 GeV hybrid width. Integrating the Primakoff Hybrid
production
differential cross section for a 280 GeV pion beam with this $\Gamma(\pi_1
\rightarrow \pi \gamma$) width gives 125-1250 $\mu$b. Second, a FNAL E272
measurement indicated (but with high uncertainty) that
$\Gamma(\pi_1\rightarrow\pi\gamma) \times BR(\pi_1\rightarrow\pi f_1)
\approx 250$ keV for a 1.6 GeV Hybrid candidate. This would be consistent
with the above maximum VDM $\Gamma(\pi_1 \rightarrow \pi \gamma$) estimate
for BR($\pi_1 \rightarrow \pi f_1$)  = 33\%.  With a total $\pi$ inelastic
cross section per Pb nucleus of 0.8 barn, the Primakoff Hybrid production
event rate R (events per interaction) is then R= 1.6-16 $\times$
10$^{-4}$. 

In four months of running, we obtain 1.4 $\times$ 10$^{13}$ beam pions. 
With a 1\% interaction length target, we obtain 1.4 $\times$ 10$^{11}$
interactions. Therefore, one obtains 2.2-22$\times$10$^{7}$ Hybrid
Primakoff events at 100\% efficiency.  We assume now a 50\% accelerator
operation efficiency. We also estimate a global 10\% average detection
efficiency over all decay channels for tracking, $\gamma$ detection,
$\eta$ acceptance and identification, trigger acceptance, global geometric
acceptance, and event reconstruction efficiency.  All these effects give a
global efficiency of 5\%. Therefore, we may expect to observe a total of
1.1-11$\times$10$^{6}$ Hybrid decays in all decay channels. For example,
following the Close and Page predictions, we may expect 24\% in $\pi f_1$,
2-8\% in $\pi \rho$, 67\% in b$_1\pi$, 0-4\% in $\eta\pi$, 0-4\% in
$\eta'\pi$, etc.

For 2, 2.5, 3.0 GeV mass Hybrids, the number of useful events
decreases by
factors of 6, 25, and 100, respectively. But even in these cases, assuming
again a global 5\% efficiency, that represents very interesting
potential samples of 1.8-18$\times$10$^{5}$, 4.4-44$\times$10$^{4}$, and
1.1-11$\times$10$^{4}$
Hybrid meson detected events, with masses 2, 2.5,  and 3 GeV respectively.

COMPASS can study hybrid meson candidates near 1.4, 1.6, 1.9 GeV produced
by the Primakoff and Diffractive processes. COMPASS should also be
sensitive to pionic hybrids in the 2-3 GeV mass range. We may obtain
superior statistics for hybrid states if they exist, and via a different
production mechanism, without possible complication by hadronic final
state interactions. We may also get important data on the different decay
modes for this state. The observation of these/other hybrids in different
decay modes and in a different experiment would constitute the next
important step following the evidence so far reported.

COMPASS provides a unique opportunity to investigate QCD hybrid exotics,
via diffractive and Primakoff production. Taking into account the very
high beam intensity, fast data acquisition, high acceptance and good
resolution of the COMPASS setup, one can expect from COMPASS the highest
statistics and a `systematics-free' data sample that includes many tests
to control possible systematic errors. Intercomparisons between COMPASS
and other experiments with complementary methodologies should allow fast
progress on understanding hybrid meson structure, and on fixing the
systematic uncertainties. 

\begin{center}
{\bf  3. Plan of Operation:\\}
\end{center}

\noindent

COMPASS studies hybrid meson structure via the scattering of high energy
pions from ``photon"  and ``Pomeron" targets.  We use 100-280 GeV beams
($\mu$, $\pi$)  and magnetic spectrometers and calorimeters to measure the
complete kinematics of pion-photon and pion-Pomeron hybrid production
reactions. Initial COMPASS set up runs are scheduled for Summer 2000. 
COMPASS will then study ``proton spin" using a muon beam to measure the
gluon polarization, and Hybrid meson structure using a pion beam. These
programs will benefit from high statistics, excellent beam focusing and
momentum analysis, and dedicated low background runs. We will analyze pion
and muon test run data, we will carry out simulation studies, and we will
plan and analyze dedicated COMPASS runs to maximize quality results. 

More accurate physics and trigger simulations are required using the new
C-programmed COMPASS GEANT package. This code should include the updated
experimental setup, trigger and the DAQ schemes, accurate magnetic field
mapping, and event reconstruction. We work on this package, incorporating
a Primakoff hybrid meson structure event generator. We develop COMPASS
event reconstruction algorithms, and test them on GEANT simulated events. 

For the COMPASS Hybrid meson structure effort, we need to plan, construct
and implement all hardware and software for the trigger. We will prepare
the COMPASS pion-photon Primakoff trigger system by the following phases:
(1) continue to investigate the hodoscope matrix energy loss and
multiplicity 3 triggers, mentioned above, via MC simulations, (2)  refine
our MC trigger simulations using COMGEANT, (3) construct the trigger
hardware, including an upgrade of the existing CEDARS Cherenkov beam PID
detector, scintillation hodoscopes, fast signal summing circuitry,
mechanical supports, etc., (5) install the system at CERN, (6) set up the
trigger detectors and electronics in the COMPASS muon beam, (7)  take
preliminary data with muons, writing event reconstruction algorithms, and
checking trigger performance, (8) use it for running the COMPASS hybrid
meson structure 
experiment with pion beams.

\vspace{0.5cm}
\begin{center}
{\bf  4. Acknowledgments:\\}
\end{center}

This work was supported by the Israel Science Foundation founded by the
Israel Academy Sciences and Humanities, Jerusalem, Israel; and the US
Department of Energy.  Thanks are due to F. Bradamante, S. Paul, L. 
Landsberg, T.  Ferbel, D. Casey, G. Mallot, H.-W. Siebert, V. Steiner, D. 
von Harrach, A.  Bravar, M.  Faessler, A.  Singovski, J. Pochodzalla, A. 
Olchevski, S. Prakhov, Y. Khokhlov, H. Willutzki, and T. Walcher for
valuable discussions. 

\newpage 
\brlist

\bibitem{compass} F. Bradamante, S. Paul et al., CERN Proposal COMPASS,\\
http://wwwcompass.cern.ch/, CERN/SPSLC 96-14, SPSC/P 297;  \\
Oct. 1999 COMPASS
collaboration meeting, summarized at\\
http://www.e18.physik.tu-muenchen.de/compass/compass-week/talks/.

\brf{cd2} M. A. Moinester, V. Steiner, Pion and Kaon Polarizabilities and
Radiative Transitions, in Proceedings of the `Chiral Dynamics Workshop:
Theory and Experiment,' U. Mainz, Sept. 1997, Eds. A. Bernstein, D.
Drechsel, and T. Walcher, Springer-Verlag, 1998, HEP-EX/9801008. 

\bibitem{hadron1} M. A. Moinester and V. Steiner, Proc., JINR (Dubna) CERN
COMPASS Summer School, Charles U., Prague, Czech Republic, Aug. 1997, Eds.
M. Chavleishvili and M. Finger, HEP-EX/9801011.

\bibitem{bormio} M. A. Moinester, V. Steiner, S. Prakhov, 
Hadron-Photon Interactions in COMPASS, Proceedings XXXVII International
Winter Meeting on Nuclear Physics, Bormio, Italy, Jan. 1999,
hep-ph/9910039

\bibitem {tb} T. Barnes, 
PHOTOPRODUCTION OF HYBRID MESONS,
Contribution to 3rd international Conference in Quark Confinement
and Hadron Spectrum (Confinement III), Newport News, VA, June 1998, 
nucl-th/9907020 

\bibitem {ni} N. Isgur 
FLUX TUBE ZERO POINT MOTION, HADRONIC CHARGE RADII, AND HYBRID MESON
PRODUCTION CROSS-SECTIONS,
Phys. Rev. D60 (1999)114016, 
hep-ph/9904494 

\bibitem {fluxtb} F. Close, P. Page, Nucl. Phys. B443 (1995) 233

\bibitem {nib98} N. Isgur, SPECTROSCOPY - AN INTRODUCTION AND OVERVIEW,
JLAB-THY-99-03-A, Feb 1999, in Proceedings of the 8th International
Conference on the Structure of Baryons (Baryons 98), Bonn, Germany, Sept.
1998, Eds. D. W. Menze, B. Metsch, World Scientific, 1999.

\bibitem {kpb98} K. Peters, Meson Spectroscopy and Exotic Quantum Numbers,
in Proceedings of the 8th International Conference on the Structure of
Baryons (Baryons 98), Bonn, Germany, Sept.  1998, Eds. D. W. Menze, B.
Metsch, World Scientific, 1999. 

\bibitem {sgjn} S. Godfrey, J. Napolitano, LIGHT MESON SPECTROSCOPY,
Rev. Mod. Phys., in press, hep-ph/9811410.

\brf{ihep1} D. Alde et at. 
Proc. of HADRON-97, BNL, August 1997. More references there.
\brf{ihep2} Yu.D.Prokoshkin and S.A.Sadovsky, Phys. At. Nucl. 58 (1995) 606.

\brf{ves} G. M. Beliadze et al., Phys. Lett. B313 (1993) 276--282;
A.Zaitsev, Proc. of HADRON-97, BNL. August 1997.

\brf{na12} D. Alde et at., Phys. Lett. B205 (1988) 397.

\brf{kek} H. Aoyagi et al., Phys. Lett. B314 (1993) 246--254.

\brf{E852} D. R. Thompson et al., Phys. Rev. Lett. 79 (1997) 1630
and BNL Press Release 97-91 
(http:$//$lemond.phy.bnl.gov/$~$e852/home\_e852.html);
G. S. Adams et al., Phys. Rev. Lett. 81 (1998) 5760.

\brf{CBa} A. Abele {\it et al.}, Phys. Lett. B423, 175 (1998).
\brf{CBb} A. Abele {\it et al.}, Phys. Lett. B446, 349 (1999).

\brf{E852rev} G. S. Adams et al., Phys. Rev. D (to be published).

\bibitem {whs} V. Dorofeev, for the VES collaboration, NEW RESULTS FROM
VES, Workshop on Hadron Spectroscopy (WHS 99), Rome, Italy, March 1999,
hep-ex/9905002

\brf{ziel} M. Zielinski et al., Phys. Rev. Lett. 52 (1984) 1195.

\brf{ciha} S.~Cihangir et al., Phys. Lett. 117B (1982) 119,
Ibid, p.123; Phys. Rev. Lett. 51 (1983) 1

\brf{jens} T. Jensen {\em et al.}, Phys. Rev. 27D, 26 (1983).

\bibitem {zihy} M. Zielinski et al., Zeit. Phys. C, Particles and Fields
31, 545 (1986);  Zeit. Phys. C, Particles and Fields 34, 255 (1987); SELEX
reports.  

\brf{mchy} D. Casey, M. A. Moinester, S. U. Chung, T. Ferbel, HYBRID,
Monte Carlo Event Generator for Hybrid Meson Primakoff Production, and
Decay, 1999. 

\brf{jens} T. Jensen {\em et al.}, Phys. Rev. 27D, 26 (1983).

\brf{pigam} M. Zielinski {\em et al.}, Phys. Rev. 29D, 2633 (1984).

\bibitem {h741} V. Steiner, M. A. Moinester, J. Russ et al.,
SELEX/E781 Laser Light Distribution System, Fermilab E781 H-notes 688,
712, 734, 741, 758, 770, 776, 777, available on WWW at
http://fn781a.fnal.gov/ and http://vsnhd1.cern.ch/~murraym

\bibitem {daq} COMPASS course to future computing (21st-century data
handling arrives), Cern Courier 39 (1999) 22.

\erlist
\end{document}